\documentclass[a4paper]{PoS}
\usepackage{amsmath}
\usepackage{mathtools}
\usepackage{tikz}
\usepackage{multirow}
\usepackage{scalefnt}

\newcommand{\pth}{p_{\bot}}

\newcommand{\mh}{m_h}

\newcommand{\largeb}{large-$b$}

\definecolor{darkgreen}{rgb}{0.2,0.6,0.2}
\definecolor{darkblue}{rgb}{0.2,0.2,0.6}
\definecolor{darkred}{rgb}{0.6,0.1,0.1}

\tikzset{
  double arrow/.style args={#1 colored by #2 and #3}{
    -stealth,line width=#1,#2, 
    postaction={draw,-stealth,#3,line width=(#1)/3,
                shorten <=(#1)/3,shorten >=2*(#1)/3}, 
  }
}

\tikzset{
  ashadow/.style={opacity=.25, shadow xshift=0.07, shadow yshift=-0.07},
}

\title{Matching uncertainties in the prediction of the Higgs boson transverse momentum in the SM and beyond}

\ShortTitle{}

\author{\speaker{Emanuele Bagnaschi}\\
        Deutsches Elektronen-Synchrotron (DESY), Notkestra{\ss}e 85, D-22607 Hamburg, Germany\\
        E-mail: \email{emanuele.bagnaschi@desy.de}}

\abstract{We present the results of our recent study~\cite{Bagnaschi:2015bop} of the theoretical uncertainties that affect the predictions for the Higgs-boson transverse-momentum in gluon fusion when fixed- and all-order 
results are matched. Our investigation consists of a twofold analysis: first we present a detailed comparison of two recently introduced prescriptions for the determination of the matching scale~\cite{Harlander:2014uea,Bagnaschi:2015qta}, then we apply the results of these methods to three widely used matching frameworks, namely the {\tt aMC@NLO} and {\tt POWHEG} Monte Carlo approaches and
analytic resummation. The results of our study are applied to the production of the SM Higgs boson and of the neutral Higgs bosons
of the Two-Higgs-Doublet Model in a variety of scenarios.}

\FullConference{Fourth Annual Large Hadron Collider Physics\\
		13-18 June 2016\\
		Lund, Sweden}
\dedicated{DESY 16-173}
\begin{document}

\section{Introduction}

After the discovery of the Higgs boson at the LHC~\cite{Aad:2012tfa,Chatrchyan:2012xdj}, a quest has started to characterize the newly found particle. Its properties have already been a topic of studies of many articles whose aim was to understand the compatibility
with the Standard Model (SM) predictions. So far these studies have relied on the precise predictions available
for production and decay rates. Only in the last year the experimental measurements for differential observables have been published~\cite{Aad:2015lha,Khachatryan:2015rxa}.
As already stressed by various authors~\cite{Langenegger:2006wu,Brein:2007da,Bagnaschi:2011tu,Harlander:2013oja,Grojean:2013nya,Azatov:2013xha,Banfi:2013yoa,Dawson:2014ora,Langenegger:2015lra} the Higgs-transverse momentum ($\pth$) distribution opens the possibility of probing the loop dynamic of the gluon fusion process. This observable is therefore sensible to modifications of the Yukawa couplings and/or to the presence of new states beyond the SM ones.
The opportunity of exploiting the $\pth$ measurement is becoming more and more interesting as the LHC accumulates a new wealth of data during the second run of operations.
At the same time, when considering models of new physics with enlarged Higgs sectors, an accurate description of the transverse momentum, which can show sensible deviations from the SM prediction, is required.

To properly describe the transverse momentum distribution, whose fixed-order prediction is logarithmically divergent in the limit $\pth \to 0$,
one needs to resum the terms enhanced by powers of $\log(\pth/\mh)$ to all orders in $\alpha_s$, where $\mh$ is the mass of the scalar resonance.
This resummation is usually performed either analytically or algorithmically.
Due to the theoretical formalism on which it is based, the resummation procedure is strictly valid only in the limit of collinear emissions and therefore, for gluon fusion,
in the limit of zero transverse momentum of the Higgs boson. Therefore, to properly describe the whole $\pth$-spectrum, a matching
between the fixed order and the resummed results is needed. Particular care has to be taken to avoid any kind of double counting.
This has been achieved in various frameworks, both analytic~\cite{Collins:1984kg,Bozzi:2005wk,Mantry:2009qz,Becher:2012yn} and numeric~\cite{Frixione:2002ik,Nason:2004rx}. Common to all the approaches is the introduction
of a new unphysical scale, which we will subsequently denote as the \emph{matching scale} ($\mu$), whose role is to define
the transition region between the fixed- and the all-order results. The dependence of the matched result on this scale is of higher logarithmic order,
however a careless choice can ruin the perturbative convergence of the result.
This is especially true for those processes that are characterized by more than one scale, as the one that we are considering in our study, as we will see in the next sections.

\section{The Higgs transverse momentum as a multiscale problem}

Higgs production in gluon fusion has been originally studied in the so-called Heavy Quark Effective Field Theory (HQEFT) obtained in the limit where the top-quark mass is taken to be very large compared to Higgs boson mass. The use of the HQEFT has the advantage of  reducing what is a one-loop LO process to a tree level one, with a sensible decrease of the complexity of the computation.
Under this approximation, the total cross section has been computed up to $\mathrm{N}^3\mathrm{LO}$, while differential computations are available up to NNLO (see ref.~\cite{Dittmaier:2011ti,Dittmaier:2012vm,Heinemeyer:2013tqa} for a complete reference of all the results available). 
However, being this an effective description, it is valid only if we are not probing mass scales that are equal or larger than the top quark mass. Therefore, the HQEFT is a description limited to Higgs boson masses smaller than the top quark mass, for what concerns the total cross section, and to $\pth$ less than top mass for the transverse momentum distribution. Moreover, it neglects completely the contribution coming from diagrams where the coupling of the Higgs to the gluons is mediated by a loop of light quarks. The latter are important for precision predictions in the SM and fundamental for Beyond-Standard-Model (BSM) Higgs boson production, where it can happen for specific regions of the BSM parameter space that the dominant contribution to the cross section is from bottom quark diagrams, oppositely to the SM. 
Therefore, to properly describe the Higgs boson transverse momentum, we need to perform the computation in the full theory, being it either the SM or a BSM model as the THDM or the MSSM.
In the case of the SM and the THDM, complete computations are available up to NLO. In the case of the MSSM, even the full NLO result is not known analytically.

Restricting ourselves, for simplicity to a THDM-like scenario, the description of the Higgs transverse momentum in gluon fusion is characterized by four mass scales:
the Higgs boson mass; the top quark mass; the bottom quark mass; the $p_{\bot}$ of the radiated parton.
All these physical scales and their non-trivial interplay have to be taken into account in the choice of the matching scale.


To follow this requirement, it was first proposed by Grazzini et al.~\cite{Grazzini:2013mca} to split the complete squared matrix element into components
that are characterized just by a single mass scale or by a specific combination thereof. Originally this split was into two parts, the top contribution
and the bottom plus the interference terms. Here we follow more recent developments where the amplitude is divided into three terms~\cite{Harlander:2014uea,Bagnaschi:2015qta}.
To achieve this, we rewrite the full amplitude as
\begin{align}
| \mathcal{M} (t+b) |^2 = \overbrace{| \mathcal{M} (t) |^2}^{\mathrm{only~top}} + \overbrace{| \mathcal{M} (b) |^2}^{\mathrm{only~bottom}} + \underbrace{\left( | \mathcal{M} (t+b) |^2 - | \mathcal{M} (t) |^2 - | \mathcal{M} (b) |^2 \right)}_{\mathrm{interference}} \,,
  \label{eq:decomp}
\end{align}
then compute separately the matched prediction for each contribution, with a different matching scale each, and finally sum all the three components together.
We note that, while the decomposition given in eq.~(\ref{eq:decomp}) is a trivial identity at the level of the total cross section,
because the latter is independent on the matching scale, it yields a modified shape for the transverse momentum distribution with respect to the one obtained
using a single scale for the full amplitude. Our master formula for the best prediction of the Higgs boson transverse momentum distribution is therefore given by
\begin{align}
  \frac{d \sigma}{d p_{\bot}} = \frac{d \sigma_t}{d p_{\bot}} \bigg|_{\mu_t} + \frac{d \sigma_b}{d p_{\bot}} \bigg|_{\mu_b} + \frac{d \sigma_{int}}{d p_{\bot}} \bigg|_{\mu_{\mathrm{int}}}\,,
  \label{eq:decomppt}
\end{align}
where with $\mu_t$, $\mu_b$ and $\mu_{\mathrm{int}}$ we have denoted respectively the matching scale for the top and bottom contributions and the interference term.

\subsection{Matching scale determination}

\begin{figure}
  \centering
  \includegraphics[width=0.43\textwidth]{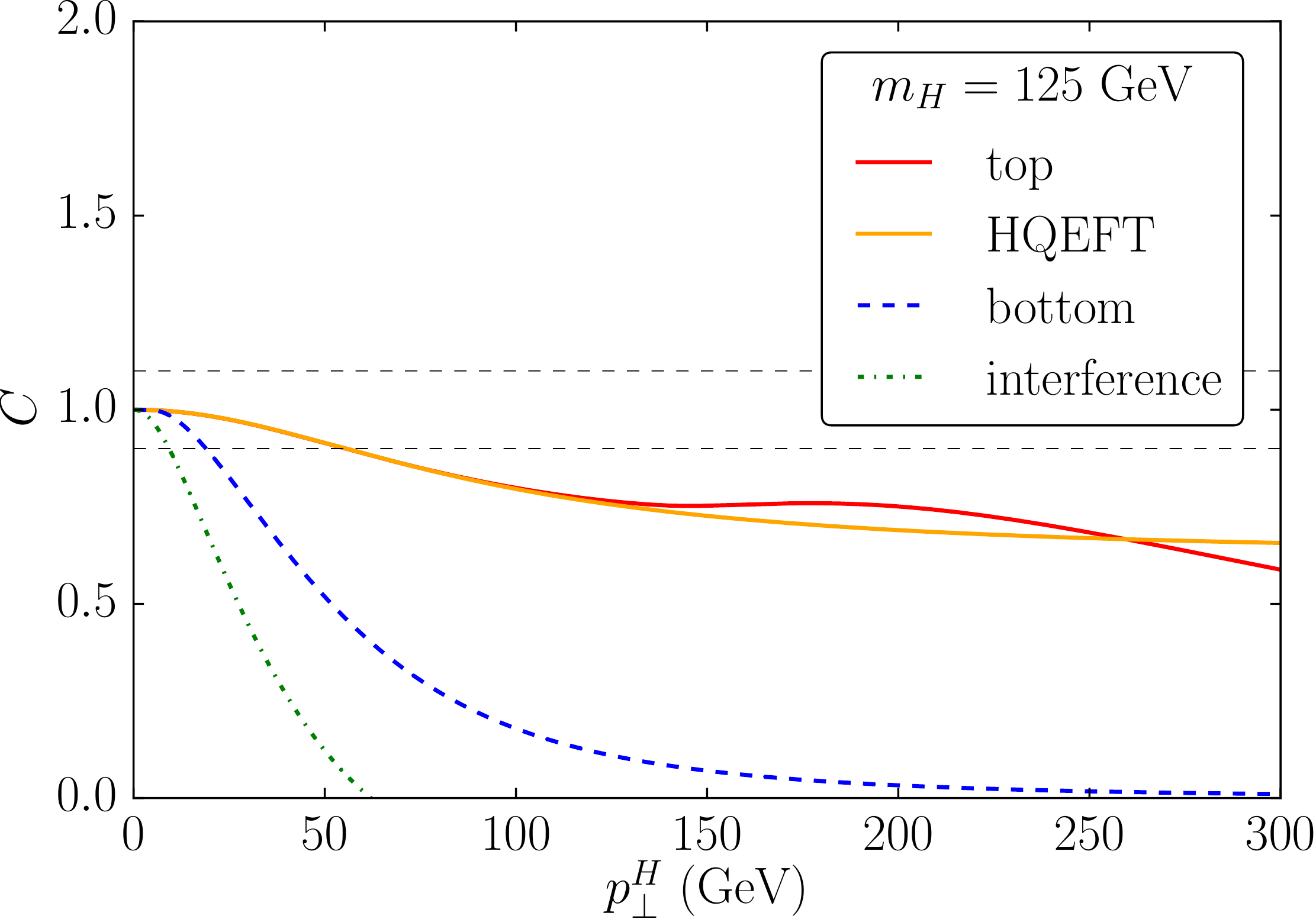}
  \includegraphics[width=0.43\textwidth]{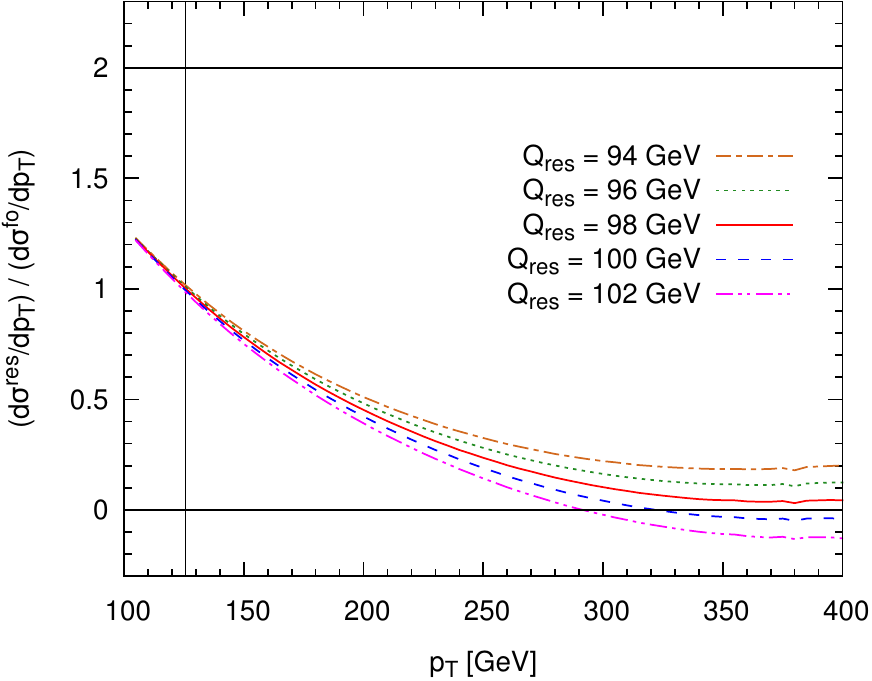}
  \caption{On the left, ratio ($C$) between the full squared matrix amplitude and the collinear limit of the gluon initiated subprocess for the top, bottom and interference terms as well as for the HQEFT computation, as a function of the Higgs transverse momentum; the dashed lines indicate a deviation of $\pm 10\%$ from one. On the right, ratio between the matched and the fixed order predictions at the hadronic level, for the top contribution and for different values of matching scale.
    The latter is here denoted by $\mathrm{Q}_{\mathrm{res}}$ because the results shown are obtained in the framework of AR.}
  \label{fig:scaledetermination}
\end{figure}

In our work we have compared two prescriptions that recently appeared in the literature\footnote{In the context of SCET, though only in the HQEFT, a detailed study on the problem of scale determination has been published in ref.~\cite{Ligeti:2008ac,Abbate:2010xh,Berger:2010xi}.}:
the partonic analysis published in ref.~\cite{Bagnaschi:2015qta} (BV) and the hadronic analysis in ref.~\cite{Harlander:2014uea} (HMW). 

The BV prescription is based upon the observation that the resummation formalism relies on the factorization of the squared matrix element in the limit of soft and/or collinear emissions (see also ref.~\cite{Banfi:2013eda}). 
In detail, the accuracy of the collinear approximation of the gluon fusion process, as a function of $\pth$, is evaluated at the partonic level, separately for each term in eq.~(\ref{eq:decomp}).
If the approximation is violated by a well-defined threshold, here chosen to be $10\%$, for a given value of $\pth$, then the latter is taken to be the matching scale to be used in the matched computation (see the left plot of fig.~\ref{fig:scaledetermination}).
The procedure is applied separately to the gluon-gluon and the quark-gluon subprocesses. The results are then averaged with a differential-weight that keeps into account
that the two channels contribute with varying proportions at different $\pth$ due to the distinct behavior of the quark and gluon PDFs.

The HMW method follows from two principles: for $\pth \gtrsim \mh$ the spectrum is correctly described by fixed-order perturbation theory and therefore the latter should be the definitive prediction
in this range; one would like to have an all-order result for a $\pth$-range as large as possible. Practically, this translates into the definition of a scale $Q^{max}_{res}$
as the maximum scale for which the resummed distribution is within the interval $[0,2] \cdot [d\sigma^{fNLO}/d\pth]$ for $\pth \geq \mh$. The matching scale $\mu$ is then taken to be equal
to half $Q^{max}_{res}$ (see the right plot of fig.~\ref{fig:scaledetermination}).

We stress that, in both cases, the scales are independent on the Yukawa coupling of the quark to the Higgs.

\subsection{Matching scale comparison}

\begin{figure}[t]
  \centering
  \includegraphics[height=0.28\textheight]{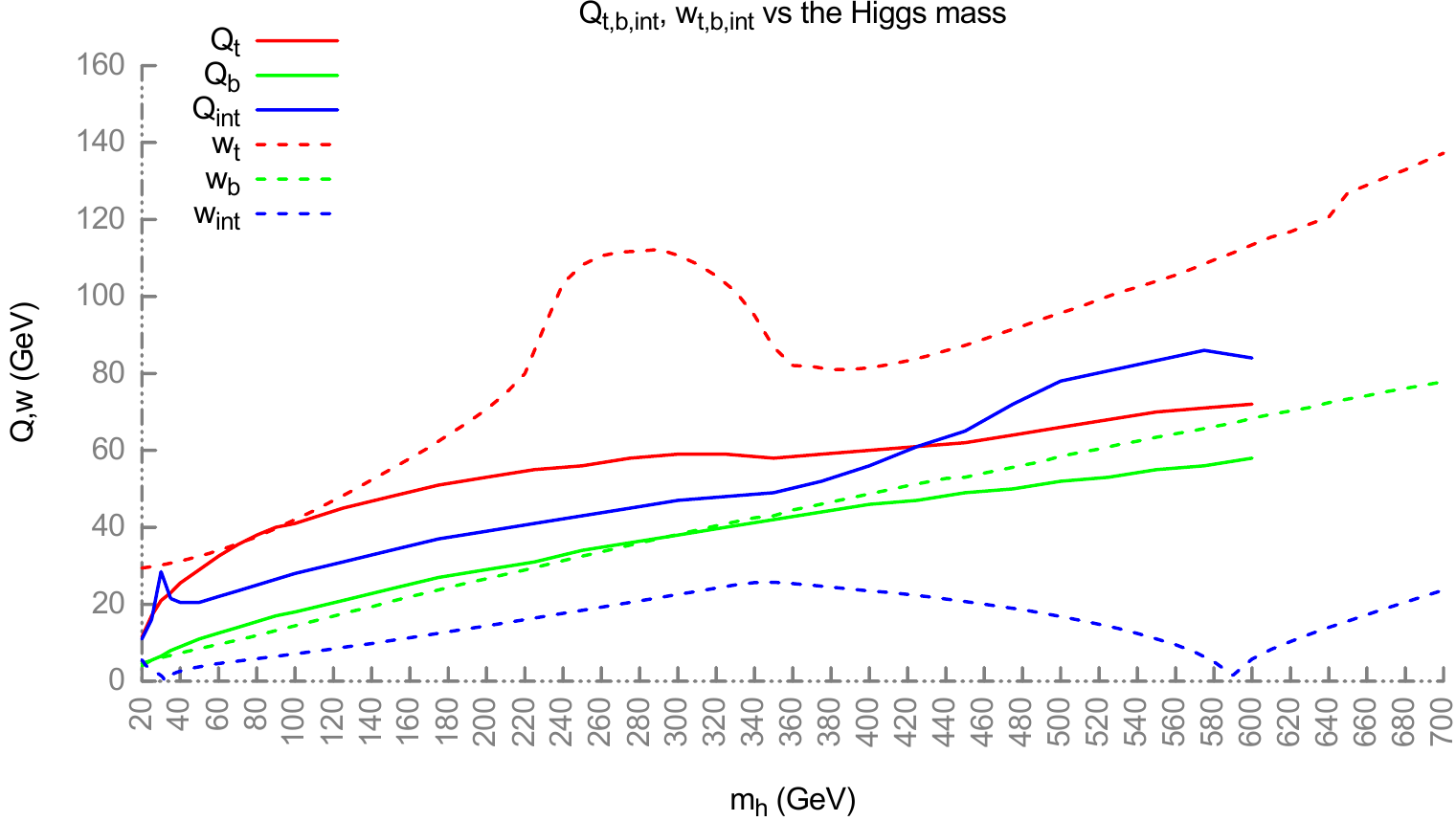}
  \caption{Comparison of the matching scales, for the top (red) and bottom (green) contributions and for the interference term (blue), obtained with either the BV ($w_i$, dashed lines) or the HMW prescription ($Q_i$, solid lines) as a function of the Higgs mass.}
  \label{fig:scalecomparison}
\end{figure}

In fig.~\ref{fig:scalecomparison} we show the two scale sets as a function of the Higgs boson mass, for the scalar case. Due to the different assumptions on which the two procedures are based, it is not surprising that the numerical values
are different. Indeed the BV prescription is sensitive to the behavior of the transverse spectrum in the low-$\pth$ region, while the HMW method is designed around the high-$\pth$ tail. What we observe
is a moderate agreement for the top contribution (with only the BV scale showing a sensitivity to the $2 \cdot m_{\mathrm{top}}$ threshold) and a very good agreement for the bottom one. Instead, for the interference term the two prescriptions can differ by a large amount and specifically this happens when the LO term
is much smaller than the NLO one. Indeed, in this case we have that the resummed contribution, being proportional\footnote{Apart from corrections due to the virtual contributions that are small compared to the total cross section.} to the LO term, is small and the Higgs transverse momentum will be given almost completely by the hard emission from the NLO term. Then the collinear approximation will fail for any value of $\pth > 0$ and therefore the BV scale will vanish; on the contrary, because the matched curve is almost identical to the fixed order one, for every value of the resummation scale, the HMW scale will tend to be very large.
As a general feature of both scale sets, we observe that, for heavy Higgs masses, the scales for all the three contributions are much smaller than the commonly used value of $\mh/2$.

\section{Simulation setup}
\label{sec:setup}

In our study of the theoretical uncertainties of the Higgs transverse momentum distribution we used three different codes: {\tt MoRe-SusHi}, which implements the Analytic Resummation (AR) procedure; {\tt gg\_H\_2HDM} from the {\tt POWHEG-BOX} framework; {\tt aMCSusHi}, based upon the {\tt aMC@NLO} framework and the {\tt SusHi}~\cite{Harlander:2012pb} amplitudes.
The uncertainty band due the matching is determined using the following prescription: given the reference values $(\mu_t,\mu_b,\mu_{\mathrm{int}})$ for the
three matching scales discussed in the previous sections, we consider all the possible combinations generated by taking half and twice these values or the reference values themselves;
for each combination we compute our prediction for the Higgs $\pth$; finally we take the envelope for each $\pth$ bin, i.e. we take the maximum and the minimum value among all the predictions. Only for AR, we follow ref.~\cite{Harlander:2014uea} and we apply an additional damping factor to the error band at large $\pth$.

In our study, besides the SM, we considered various THDM-II scenarios. Each of the latter was chosen because it is characterized by a specific feature, e.g.~a very large Yukawa coupling to one of the two quarks, whose impact on the $\pth$ distribution we want to understand. We also considered both scalar and pseudoscalar productions. In this proceeding, due to the restricted space available, we report the results only for the SM Higgs boson using BV scales and for heavy scalar production in the \largeb{} scenario with the HMW scale set. See table~1~of ref.~\cite{Bagnaschi:2015bop} for a list of all scenarios.

All the numerical results are computed for the LHC, with a a center-of-mass energy of $\sqrt{S} = 13$ TeV, using the {\tt MSTW2008nlo68cl} PDF set and the associated value of $\alpha_s(M_Z) = 0.120179$.
The renormalization and factorization scales are set to the Higgs boson mass; the quark masses are fixed at $m_{\mathrm{top}} = 172.5$ GeV and $m_b = 4.75$ GeV respectively; the Yukawa couplings
of the Higgs to quarks are renormalized in the On-Shell (OS) scheme. We used {\tt Pythia8} as our Parton Shower~\cite{Sjostrand:2007gs}.



\section{SM results}

\begin{figure}[t]
  \centering
  \includegraphics[width=0.42\textwidth]{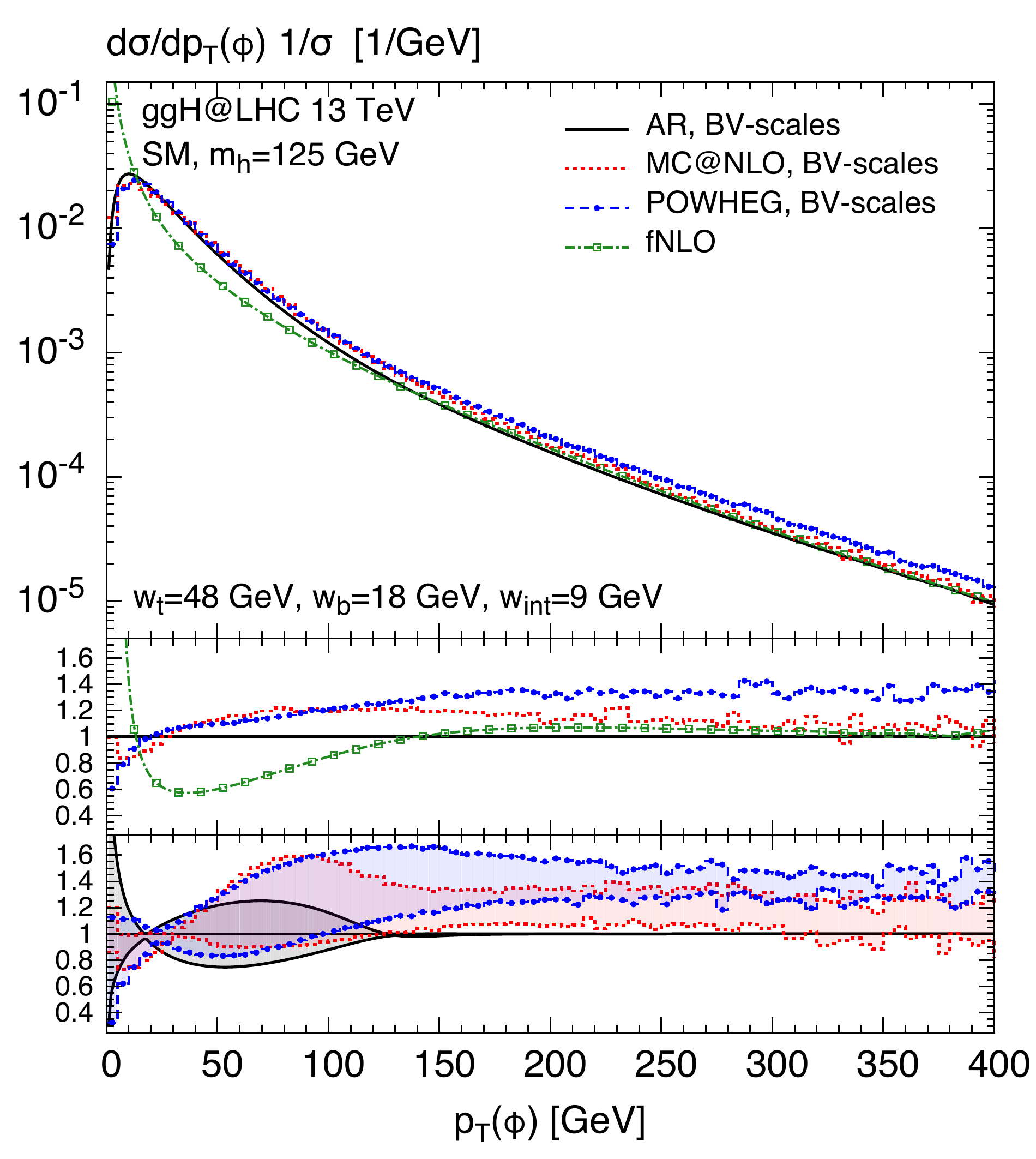}
  \includegraphics[width=0.42\textwidth]{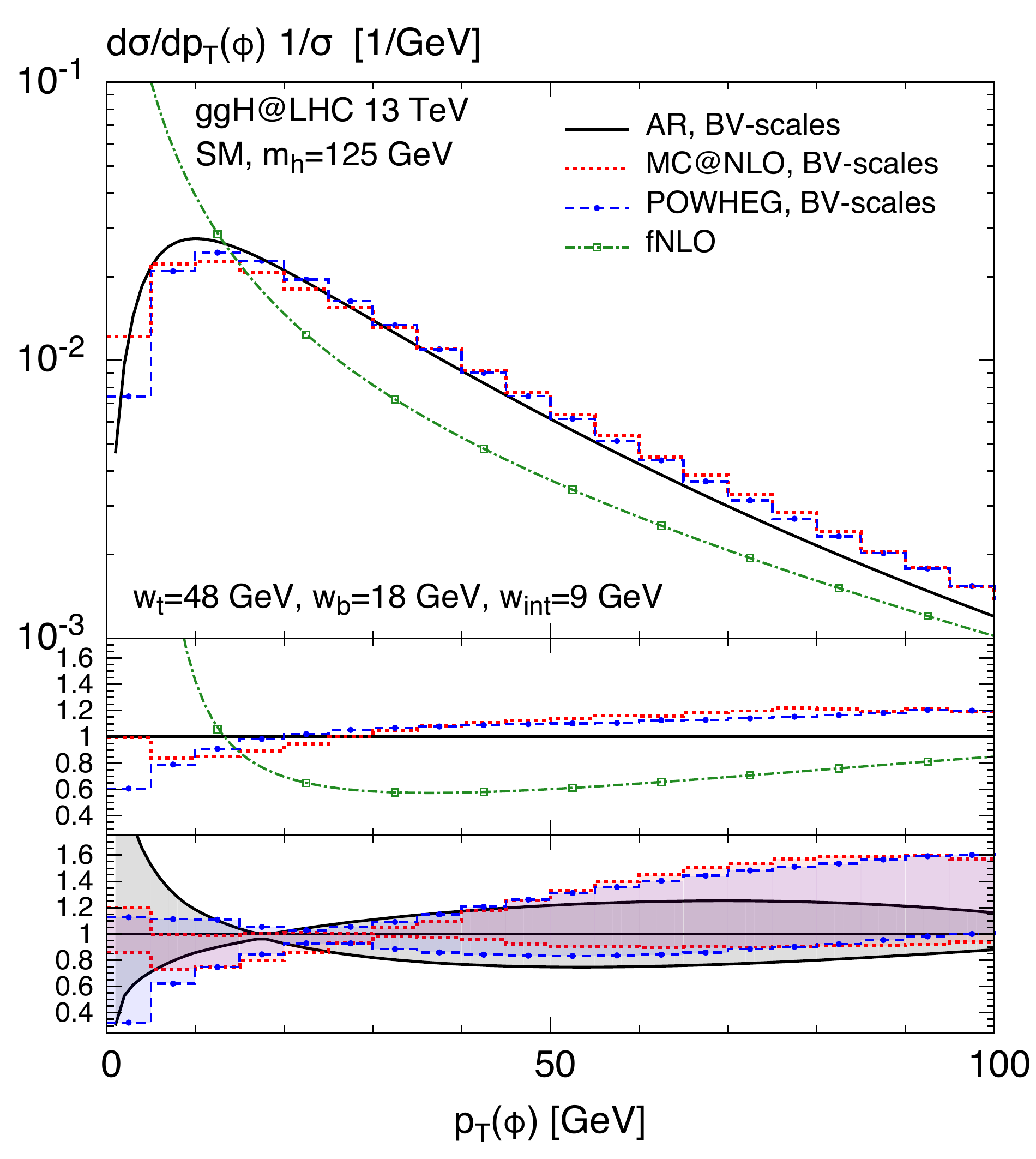}
  \caption{Shapes of the transverse momentum distribution for a SM Higgs boson of mass $m_h = 125$ GeV. We show the curves from AR
    (black, solid), {\tt MC@NLO} (red, dotted) and {\tt POWHEG} (blue, dashed overlaid by points) for the BV scale set. For reference we also show the fixed order result (green, dash-dotted with open boxes).
    On the left, we plot the spectrum up to $400$ GeV, on the right we show a zoom around the first $100$ GeV.
    In the main frame we show the absolute distributions, in the middle inset the ratio of the central curves to the analytic resummation result
    and in the bottom inset the uncertainty bands, again normalized to the AR resummation value.}
  \label{fig:smresults}
\end{figure}

In fig.~\ref{fig:smresults} we show the shape of the transverse momentum distribution (i.e. the integral of each curve
is normalized to one) obtained using the BV scale set for a SM Higgs boson of $m_h = 125$ GeV.
Qualitatively we observe that all the codes agree within their uncertainty bands, at least for $\pth < \mh$. More in detail,
we observe that the two MC event generators, {\tt POWHEG} and {\tt MC@NLO} are in excellent agreement in the
region $10 < \pth/\mathrm{GeV} < 130$, while they differ by about $20\%$ from the central AR prediction. The position of the peak is also slightly different between the MCs and AR.
Turning now our attention to the high-$\pth$ tail, we see that: the AR prediction approaches the fNLO at the level of $5\%$ above $\pth \simeq 130$ GeV;
the transition to the NLO prediction in {\tt MC@NLO} is around $\pth \simeq 180$ GeV; {\tt POWHEG}, on the other hand, always remains
$20\%$ above the fixed order result. The latter is a  general feature of the {\tt POWHEG} matching that we observe and that will be analyzed more in detail in the THDM analysis that follows.

Concerning the uncertainty bands, for $\pth > 130$ GeV  AR has no uncertainty band due to the artificial suppression given by the damping factor, as mentioned in section~\ref{sec:setup};
on the other hand, the uncertainty for the {\tt MC@NLO} prediction is of order $\pm 10\%$, while the width of the {\tt POWHEG} band decreases uniformly from $\pm 20\%$ to $\pm 10\%$.
In the intermediate region, all three codes show a bulgy structure with a maximum of $\pm 20\%$ and $\pm 35\%$ for AR and the MCs respectively and a minimum
of just a few percent above the peak position. Finally, in the small-$\pth$ region, the AR uncertainty band grows to $100\%$, the {\tt POWHEG} one to $40\%$, while
{\tt MC@NLO} shows only a moderate $\pm 15\%$ uncertainty.

\section{THDM results}

\begin{figure}[t]
  \centering
  \includegraphics[width=0.42\textwidth]{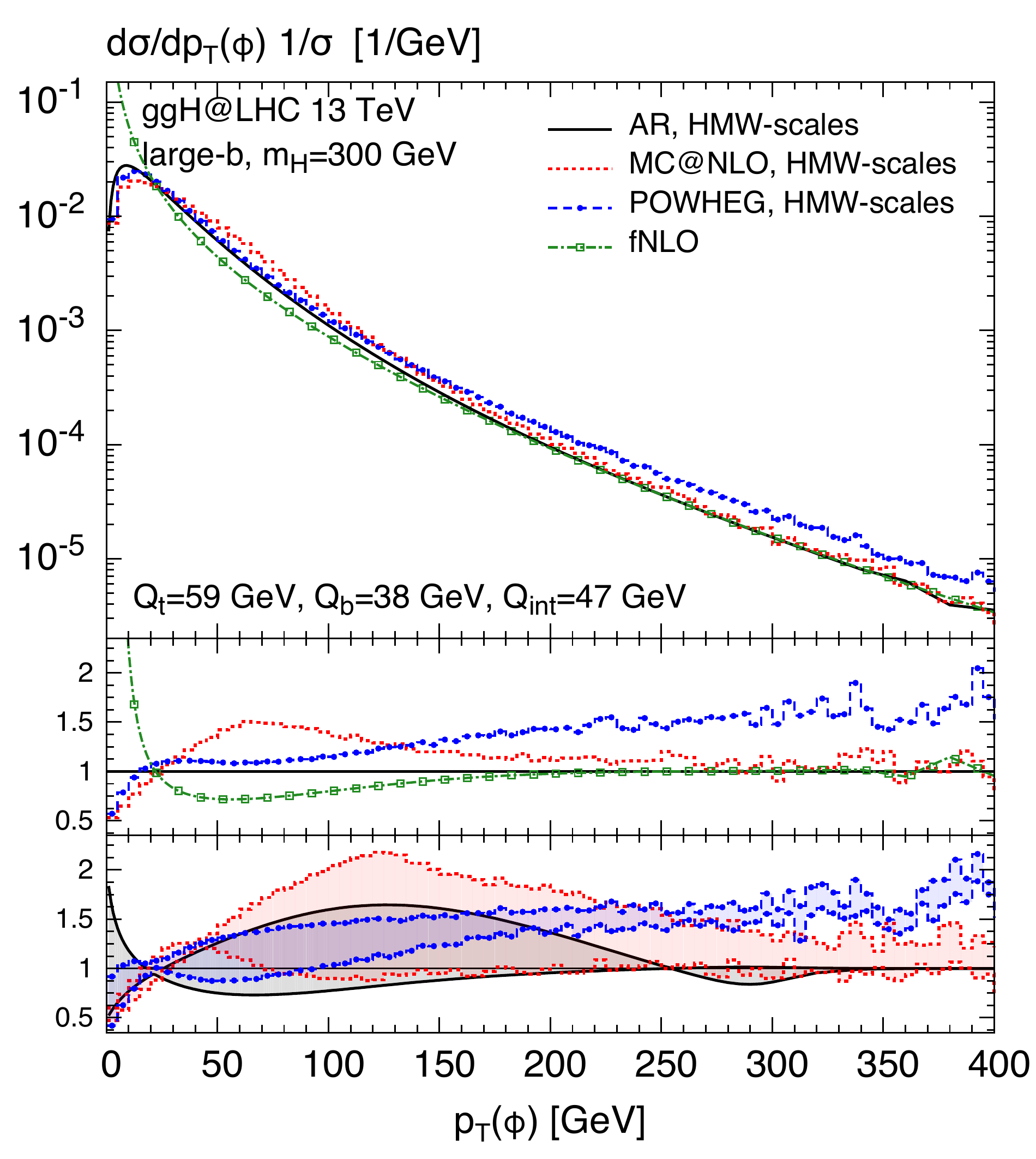}
  \includegraphics[width=0.42\textwidth]{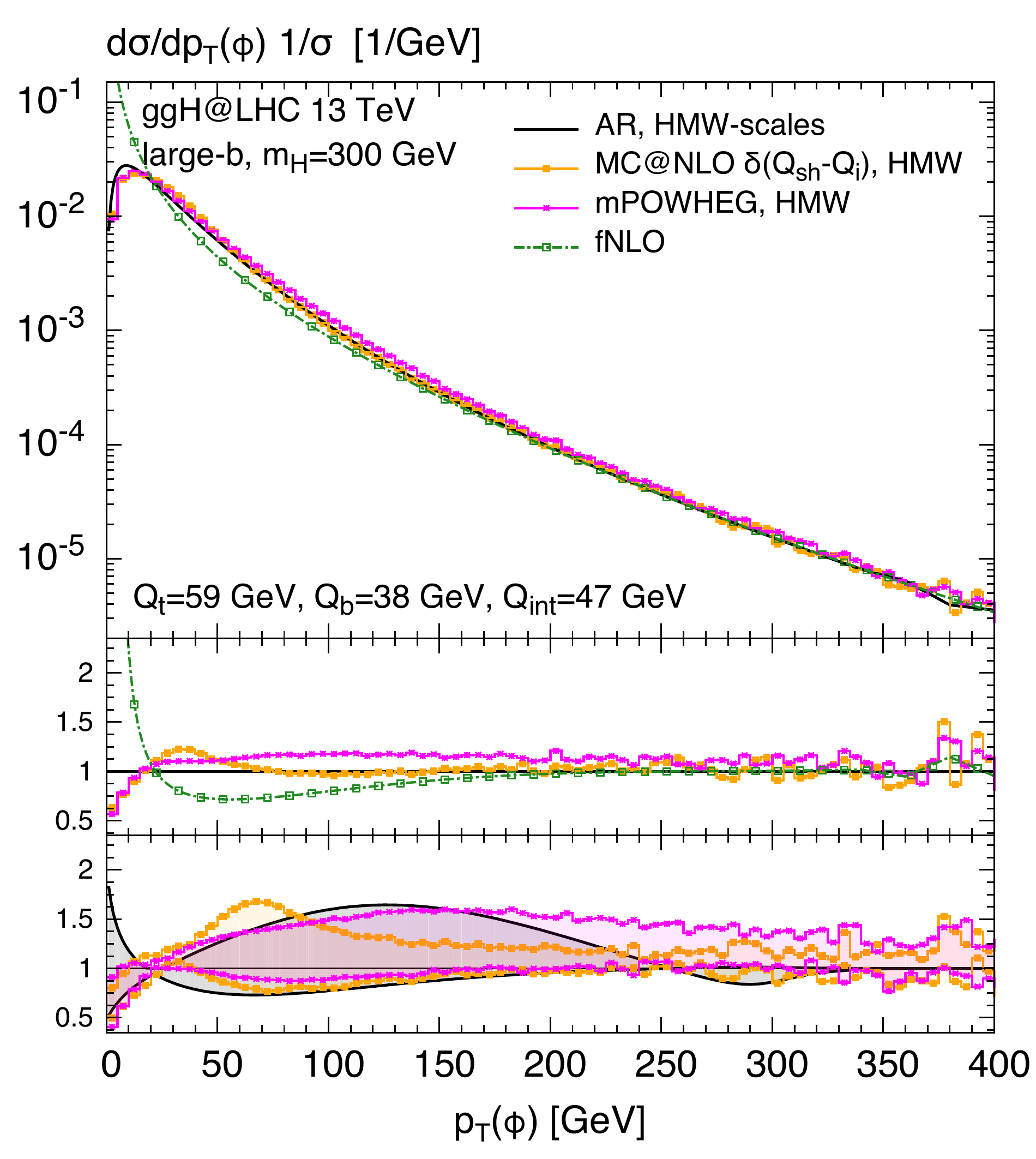}
  \caption{The plot on the left is as in fig. \protect\ref{fig:smresults} but for the production of a heavy CP-even Higgs of $m_H = 300$ GeV and for the use of the HMW scales. On the right we
  show the results with the {\tt MC@NLO} and {\tt POWHEG} curves obtained with the changes to the shower scale prescription as described in the text.}
  \label{fig:thdm}
\end{figure}

The left plot of fig.~\ref{fig:thdm} shows the same curves as the SM case discussed in the previous section. Differently from the SM, however,
the discrepancies between the three frameworks are more marked. Concerning the central curves, we observe that {\tt POWHEG} produces
a spectrum that is significantly harder, being over $50\%$ above the fNLO result for $\pth > 200$ GeV. On the other hand,
in the intermediate region between $10 \lesssim \pth/ \mathrm{GeV} \lesssim 130$, {\tt POWHEG} and AR agree within $10\%$ while the {\tt MC@NLO} curve
is substantially larger. At smaller transverse momentum, for $\pth < 30$ GeV, the two MCs have a much better agreement.

The behavior of the uncertainty bands is also quite different in the various frameworks: the {\tt MC@NLO} band blows up to $\mathcal{O}(100\%)$
around $\pth \simeq 125$ GeV; the {\tt POWHEG} band remains very small all over the whole $\pth$ range.

To understand the origin of these differences, we investigated the dependence of the MC predictions on the prescription for the shower scale $Q_{sh}$. The latter is  the scale that it is passed to the PS to be used as an upper bound for the $\pth$ of the emitted radiation.
Concerning {\tt POWHEG}, we notice that by restricting, for the class of events\footnote{These are the \emph{remnant} events. By default setting of the {\tt POWHEG-BOX} their shower scale
  is set to the transverse momentum of the emitted parton. See ref.~\cite{Alioli:2010xd} for a detailed description of the {\tt POWHEG-BOX}.}
that describe the high-$\pth$ tail, the shower scale to be at most the matching scale (either BV or HMW), we recover the fixed order result in the same way as AR and {\tt MC@NLO} do (as it can be seen from the purple curve on the right plot of fig.~\ref{fig:thdm}).
Moreover, the shape of the uncertainty band is also changed, showing now a bulge between $50$ GeV and $100$ GeV.
Relatively to {\tt MC@NLO}, we first recall that by default the shower scale is extracted from a probability distribution dependent on the LO kinematic and centered around the matching scale~\cite{Alwall:2014hca};
if we replace the default distribution with the $\delta$-function $\delta(Q_{sh}-\mu_i)$, we observe a significant change for both the central prediction and uncertainty bands, as it can be seen by the yellow curve in the right plot of fig.~\ref{fig:thdm}.
These observations lead us to conclude that for this specific scenario there is a high sensitivity not only to the numerical values of the scales but also to the specific details of the matching procedure.

\section{Conclusions}

In this talk we have presented the results of our recent study~\cite{Bagnaschi:2015bop} of the theoretical uncertainties intrinsic to the matching procedure
between fixed- and all-order results in the computation of the transverse momentum distribution of the Higgs boson in gluon fusion.
Specifically for this process, which involves different mass scales, even the choice of the central values for the matching scales has become a
matter of debate.
In this context, we performed a thorough analysis of the predictions obtained using three different matching frameworks (analytic resummation, {\tt POWHEG} 
and {\tt MC@NLO}) and two different prescriptions for the determination of the matching scales (BV or HMW). 
Our comparison was twofold: first we addressed the issue of the determination of the central value for the matching scale for
the top and bottom contributions and for the interference term, by providing a qualitative and quantitative comparison of the BV and HMW approaches;
then we compared the results for the shape of the $\pth$ distribution obtained with different scale-choices and frameworks.

We have found that the prediction of the Higgs transverse momentum is affected by uncertainties\footnote{In this study we have limited ourselves to the matching uncertainty, however the latter should always be combined with the fixed-order perturbative uncertainty, usually estimated through the variation of the renormalization and factorization scale.} up to several tens of percent, depending on the scenario,
the $\pth$ value and the framework under consideration. 
In the low-$\pth$ region we find reasonable agreement between the different codes, although
AR usually shows a much softer spectrum. However, in all the three frameworks and especially for AR, the error bands grow in this region, therefore
providing compatibility between the different results.
In the intermediate region, we find non-trivial differences between the three frameworks, which are more pronounced in the bottom dominated scenarios. In the latter case, we also find a large dependence on the specific details of the matching formulation inside each framework.
In the large-$\pth$ tail, where technically all the codes have LO accuracy~\footnote{In the SM case, new developments are available which provide NLO-QCD accuracy in the description of the Higgs $\pth$, see refs.~\cite{Hamilton:2013fea,Hoche:2014dla,Hamilton:2015nsa,Alioli:2013hqa,Frederix:2016cnl}.}, we find that {\tt POWHEG} systematically predicts a harder spectrum
than {\tt MC@NLO} and AR. The latter are instead softer and compatible with the fNLO result. We identified one source of this difference in the
prescription used to define the allowed phase space for radiation emission by the PS. Restricting the phase space, as it happens already in the {\tt MC@NLO} framework, allows also {\tt POWHEG} to approach the fNLO at high-$\pth$, as we have shown with a dedicated analysis.


\end{document}